# New Pedagogy for Using Internet-Based Teaching Tools in Physics Course

*David Toback*[1,§], *Andreas Mershin*[1,2,#], *Irina Novikova*[1,3,♦]

[1] *Department of Physics, Texas A&M University, College Station, TX 77840-4242*

[2] *Center for Biological Engineering, Massachusetts Institute of Technology,
77 Massachusetts Ave., Cambridge, MA 02139, USA*

[3] *Harvard-Smithsonian Center for Astrophysics, Cambridge, MA 02138*

Acquiring the mathematical, conceptual, and problem-solving skills required in university-level physics courses is hard work, and the average student often lacks the knowledge and study skills they need to succeed in the introductory courses. Here we propose a new pedagogical model and a straight-forwardly reproducible set of internet-based testing tools. Our work to address some of the most important student deficiencies is based on three fundamental principles: balancing skill level and challenge, providing clear goals and feedback at every stage, and allowing repetition without penalty. Our tools include an Automated Mathematics Evaluation System (AMES), a Computerized Homework Assignment Grading System (CHAGS), and a set of after-homework quizzes and mini-practice exams (QUizzes Intended to Consolidate Knowledge, or QUICK). We describe how these tools are incorporated into the course, and present some preliminary results on their effectiveness.

**Introduction**

The average college student often does not have the appropriate preparation level and study skills to succeed in introductory-level physics courses. In addition to class attendance and/or group study, students must also learn to work by themselves to develop the mathematical, conceptual, and problem-solving skills they need. While computerized homework and quiz grading programs (see, for example [1,2]) offer exciting possibilities (including the reduction of costly and time-consuming human grading) [3,4], many simply have the equivalent of having students turn in their homework online, and studies have shown this yields no performance advantage over homework turned in on paper if grading for the latter is quick, thorough and consistent [3]. We propose a new pedagogical model that takes advantage of modern information technology and helps our students where they need it most.

In this paper we describe some of the roadblocks our students encounter during their course and the steps we have taken to remove them. Our pedagogy is based on three fundamental principles: balancing skill level and challenge, providing clear goals and feedback at every stage, and allowing repetition without penalty. Using these principles we have implemented a set of internet-based testing tools which include math quizzes, homework collection and grading, homework quizzes, and mini-practice exams. Students that use these systems see where they need to concentrate their efforts, become more comfortable with the mathematics they need, and are taught powerful ways to study by themselves. We conclude with some preliminary results from using these systems in the context of a typical style lecture-based course over a number of semesters.

**The Problem and the Pedagogy**

To succeed in their first-year, calculus-based physics courses (Classical Mechanics and Electricity & Magnetism) students must develop analytical problem-solving skills. Not only must they learn to understand and use the physics concepts in a problem, they must turn the physical quantities into variables or symbols, "translate" the problem/concepts into equations, and "turn the crank" on the mathematics to find the answer (a "closed-form" solution). Our experience is that majority of our students (mainly undergraduate engineers at Texas A&M University), has little experience with, and in some cases a distinct fear of, symbol-based problems requiring closed-form solutions. Some do not even feel comfortable with such simple tasks as fraction addition or solving two equations with two unknowns.

To confront these problems, we encourage a learning model which is culturally familiar to most students in our classes and often produces many hours of intense, concentrated and passionate non-schoolwork effort: playing video games. Today's student understands a video game instinctively, and the fun comes not just from fancy graphics, but also from the quality of the interaction and the game's structure. Good games, following sound psychological principles [5], require focus, a balance of skill level and challenge, clear goals and feedback, as well as the opportunity to repeat the task at hand until it is no longer difficult. In a typical game there is a score so one can see how well one is doing, the goal is to get as high a score as possible. Often there are several levels, and each level is designed so that one can progress with significant but not too much effort. At home, it is trivial to start a new game and one plays as many games as one pleases in order to get past each level. Even then, just because one can pass a level once does not mean it is easy the next time; getting to the highest levels requires repetition until all the skills that are useful at various stages of the game are acquired and integrated simultaneously. Learning to get good at video games clearly resonates with today's youth. Those who play are often incredibly focused and engaged: they are active learners. The contrast with the typical student sitting in a physics lecture could not be more striking. We have used this video game model to develop a pedagogy which has students work on learning physics the same way they work at video games; our hope is to get at least some fraction of the dedicated effort and intensity.

We have created a set of three internet-based testing tools/quizzes that utilize our pedagogical model. They are i) an Automated Mathematics Evaluation System (AMES) which helps students hone/develop a mathematical "toolkit" needed to solve physics problems; ii) a Computerized Homework Assignment Grading System (CHAGS) which encourages solving all the assigned problems in both symbolic and numeric form and gives feedback along the way; and iii) a set of after-homework QUizzes Intended to Consolidate Knowledge (QUICK) which provide short new problems for students to solve. In each we begin with a clear definition of what it means to "get to the next level," we give them unlimited attempts to get a perfect score without penalty (changing the problem slightly on each attempt), and we give them feedback along the way. We make "winning" worth their time by rewarding them: guaranteed high scores towards their final grade. In order to pass the course we require they get a perfect 100% score on all assignments, in sequence. However, to balance this weighty requirement, and as a further incentive, we offer another video game concept, bonus points, for getting 100% on-time. While we have implemented these tools using WebCT [1], they can be straightforwardly implemented with other software packages.

## Automated Mathematics Evaluation System (AMES)

During the first week of the semester we assign a set of online mathematics quizzes that are strictly limited to the relevant pre/co-requisite math topics for the course and are designed to be finished quickly. Our intent is to remind students of the relevant calculation tools they should already have by giving them enough practice until they have (re)gained facility, and to establish a high competency level for the course[1]. Each quiz consists of ten multiple-choice problems, developed by the authors from standard homework problems, which cover what we have identified as the most common deficiencies of our students. Every attempt randomly draws from our large pool of questions and includes at least one from each of seven different areas: i) simple algebraic expressions in one variable; ii) systems of equations in two variables; iii) quadratic equations and identities; iv) geometry and trigonometry including vectors; v) fractions, numbers, exponents, powers of ten; vi) word problems and proportionalities, and vii) simple differentiation and integration (calculus is a co-requisite for the course, but many students have seen the material previously; there are instructions for those who have not yet taken calculus). Examples of questions and AMES screen shots can be found in [6].

Students have ten minutes to complete each quiz, and they must obtain a 100% on ten "separate" quizzes (although each quiz is actually drawn randomly from the same pool of questions). Each question is designed such that a well-prepared student can easily complete an entire quiz in less than five minutes. By design, the system "levels the playing field" as those already comfortable with the mathematics can quickly pass, and those needing help re-take quizzes without penalty, until they reach the same proficiency level.

Every time a quiz is attempted, students are presented with the same questions they answered incorrectly the previous time (but with the five multiple-choice answers shuffled) and fresh ones from the same section to replace the ones they answered correctly[2]. It is also important that all the seven AMES categories are represented in each ten-question quiz; in this way students become accustomed to working simultaneously with more than one concept, such as algebra and vectors, and do not learn one only to forget the other. After finishing with AMES students are more comfortable and proficient in the entire set of physics-relevant math topics.

## Computerized Homework Assignment Grading System (CHAGS)

Our course now requires students to turn in their homework online using CHAGS. In addition to the "free-tries-until-perfection methodology," what is different about CHAGS is that it is designed for turning in homework *after* students have completed the assignment with paper and pencil and produced their answers in closed form. Students receive a list of problems from the course textbook where the typical problem is stated with numerical values for given physical variables (e.g. the mass is 10 kg, the angle is $20^o$, etc.), and they are asked to find a numerical solution (e.g. what is the value of the acceleration). We have found that inexperienced students tend to start by writing their equations with the numerical values directly entered in, and they try to manipulate the equations until they get the correct numerical answer provided in a solution manual. We urge our students to use symbolic variables, such as mass = $m$, angle= $\theta$,

---

[1] We have also begun to develop an additional set of quizzes to train students to transform a word problem into a set of equations even before learning any physics concepts.

[2] We note that this important feature of AMES was originally developed on our custom-built prototype system, but is not currently available on WebCT and has been requested from the developers.

acceleration = $a$, and then solve for $a$ in terms of $m$, $\theta$, etc.

To turn in their homework after obtaining a complete set of closed-form answers, students log in to CHAGS where they expect to see a set of problems identical to the ones they just solved, but with the different numeric values of one or two of the parameters. They have only a short time to substitute the new number(s) into their formulae, recalculate and input the new numeric result, which is then checked by the system at run-time allowing for small rounding errors. The time constraint forces students to be ready with their formulae before they log in to the system and thus emphasizes the utility of obtaining closed-form solutions. We note that although input of answers in symbolic form is possible, and in some cases preferable, this option is not available in many commercial systems [7]. An advantage we have found is that numerical answers encourage students to be careful in their calculations and to perform mental "reality checks" for the magnitudes of various physical quantities which develops intuition for the realistic values for everyday things such as the speed of a bicycle or the mass of a billiard ball.

As in AMES, the number of attempts is unlimited and we require students to correctly answer all the problems, all at once, so there is no temptation to ignore harder ones or to learn a set of topics and then forget them. In contrast, the problems are always the same, but all of the parameters are changed for every problem on every attempt. [3] This effectively discourages the time-wasting trial-and-error strategies [3]. Furthermore, studies have shown that immediately providing the correct numerical response and allowing a second chance gives students the opportunity to elucidate mistakes and enhance understanding [8]. In our case students can immediately see all the correct numerical answers and use multiple attempts, if needed, to guarantee that they find the errors in their symbolic answers.

**Quizzes Intended to Consolidate Knowledge (QUICK)**

Successful completion of the homework assignment after many hours and/or attempts does not necessarily mean that a student is prepared to solve an unfamiliar problem. Often this is the ideal time to provide real feedback to the students as to whether they truly understand the material and/or to reinforce their learning with additional practice. To do this, we have implemented a short, multiple-choice quiz for each textbook chapter using the standard test banks which are readily available (and free to instructors) [9]. Each consists of two intermediate-level problems randomly drawn from a large pool[4], selected from the textbook's test bank, necessitating roughly three to five minutes per problem to solve (we give students ten minutes total). Again, students are required to get a perfect score on a quiz for each chapter and they are allowed unlimited attempts, albeit with completely new, randomly drawn problems on each attempt.

We also offer a voluntary practice tool to help students in their preparation for each traditional in-class exam. After successfully submitting all the homework assignments and completing the associated quizzes for the chapters covered in the upcoming exam, students gain access to a mini-practice exam created using the same type of quiz problems from the same chapters. For example, our first exam, which covers chapters 1, 2, and 3, the mini-practice exam includes four randomly selected problems (one from Ch. 1, one from Ch 2 and two from Ch. 3) and the students have 20 minutes to complete all problems. This tool not only provides a meaningful feedback on student readiness for the traditional exam, it teaches students to learn the material as a whole, and not just "cram" one piece at a time. It also pushes them to study for physics exams by solving many problems instead of the common "read the book or look over your homework and figure you're ready if it makes sense" approach. Even though mini-practice exams are not part of the required course work, we have found that "earning the right" to take the mini-practice exam is an effective "carrot," and we further encourage students by explicitly offering a few bonus points for scoring a 100% before each in-class exam.

---

[3] In order to minimize both student complaints and lucky guesses, and maximize the benefit of the requirement to solve many problems at once, we typically break up an assignment of 15 small problems into three separate submissions, each with a 20-minute time limit.

[4] We note that WebCT (and other programs) do not currently allow us to use the feedback from all our teaching tools to generate personalized quizzes. We look forward to the day when the software can "learn" so we can better target the individual weaknesses of each student.

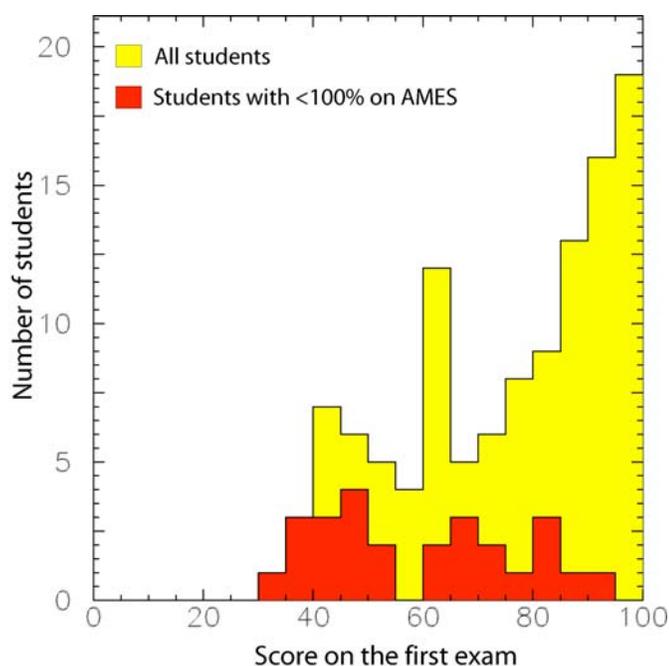

**Figure 1** A comparison of the grade distribution for the first exam for students who completed AMES (ten 100% scores) beforehand and for those who did not.

**Discussion**

We feel that our "video-game" model, which encourages repetition and facility of all relevant skills at once, has many advantages over traditional pen-and-paper homework submission and other available computer-facilitated teaching tools. Our systems have now been tested for many semesters, and Figs. 1 and 2 present statistical analysis of performance when these methods were incorporated into the classroom for the most recent semester. Both show clear correlation between performance and in-class examination scores. On the weight of this evidence alone, however, we cannot make any strong statements on the effectiveness of these tools, as smart and/or diligent students who completed our quizzes might have done well even without them. Additional, appropriately controlled quantitative studies are clearly needed to accurately evaluate the usefulness of the proposed methodologies. Our data does show that we succeed in getting the majority of the students to finish all quizzes and homework assignments on time and that most complete the voluntary mini-practice exams as a preparation for in-class tests. Perhaps most importantly, showing these figures in class provides an unambiguous message to the poorly performing students: "if you want to improve your grade, here is what you need to be doing."

We are aware that the stringent requirement that every student pass all homework assignments at 100% imposes a restriction on the difficulty level of the problems. We also note that while we find that students withdraw from our course earlier than in other courses, the final number of students who leave is roughly the same. Of those who stay, only ~3% of students do not finish the requirements when periodically reminded of them. From those there are none who would pass the course if the homework requirement were lifted. Finally, these requirements do not alter the grade distribution curve even with ~97% of the students achieving 100% scores, since this only produces a small overall shift of the mean (provided the homework/quiz portion of the grade is not larger than about ten percent). At the same time we have found "securing" 100% for at least a part of the course grade to have a profound positive impact on the morale of many students.

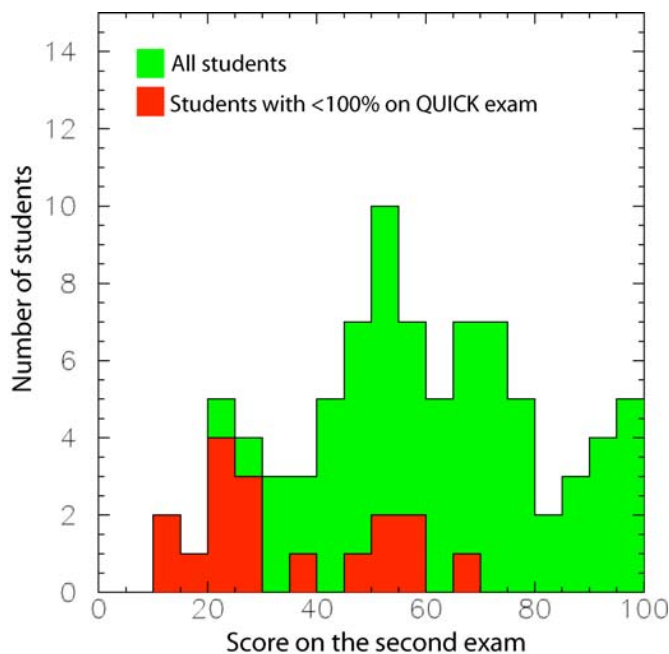

**Figure 2** A comparison of the grade distribution for the second exam for students who completed a voluntary mini-practice exam before the in-class examination and for those who did not.

*What the students thought*: There are a range of student responses and clearly not all are converts. Some students complained bitterly about being forced to spend too much time on their homework as our system, like others, does suffer from the occasional time-wasters of mistyped numbers, internet outages, broken URLs, etc[5]. Many appreciated the feedback and the opportunity to secure a high homework grade. A surprising number of students "confessed" to having developed more self-confidence in the class and said that they now understood how to study for other classes as well.

*What the teachers thought*: While some are uncomfortable with the perfection requirement and the constraints this system puts on their students, they are happy with the outcome. Prior to the existence of these systems, instructors generally heard complaints about the mathematical content of the course ("I understand the physics but can't do the math") or the difficulty of the exams ("I've never had to do a problem with variables before"). The introduction of AMES and CHAGS nearly eliminated such gripes. In addition many reported that the pervasive "symbol fear" was all but eliminated and that students no longer waste valuable class time complaining that they cannot follow the simple algebraic manipulations on the blackboard. Time and effort can now be concentrated on teaching physics concepts. If nothing more, this aspect alone makes the course more enjoyable to teach.

**Conclusions**

We have described a new pedagogy for using internet-based testing tools to develop math skills, turn in homework and take practice quizzes in introductory university-level physics courses. We have found that by using a balance between skill level and challenge, providing clear goals and feedback at every stage, and giving the student the opportunity to repeat, without penalty, until the task at hand is no longer difficult, we have addressed some of the most important deficiencies of our students. We believe this system teaches students to study in a powerful way which is new to most of them. Our preliminary results are promising and we encourage teachers elsewhere to try similar methods and possibly perform more systematic evaluations. We hope that with the pervasiveness of internet access and general computer familiarity of the average college student others will find our methods to be a useful and economical way of improving understanding and performance in their physics courses.

**Acknowledgments**

---

[5] We battle the other common problem - unclearly-worded questions, typos, incorrect solutions - by giving extra credit to the first couple students who report any broken question.

We would like to thank Wayne Saslow, Teruki Kamon, Peter McIntyre, Bob Webb, George Welch, Cathy Ezrailson and Joan Wolf for useful discussions, and Joel Walker, Matt Cervantes, Rhonda Blackburn, Jim Snell, Court Samson and Sally Yang for help with programming and other technical expertise. Financial support was provided by the Texas A&M University Department of Physics, Instructional Technology Services, and the Montague Scholarship Program at the Center for Teaching Excellence. AM was partially supported by the S.A. Onassis Public Benefit Foundation.

[#] mershin@physics.tamu.edu

[♦] inovikova@cfa.harvard.edu

[§] toback@physics.tamu.edu